\documentclass[runningheads]{llncs}

\usepackage[T1]{fontenc}
\usepackage{graphicx}
\usepackage{booktabs}
\usepackage{multirow}
\usepackage{amsmath}
\usepackage{amssymb}
\usepackage{url}
\usepackage{array}
\usepackage{float}
\newcolumntype{L}[1]{>{\raggedright\arraybackslash}p{#1}}
\title{Lip-Focused Audio-Visual Consistency Detects Speech-Content Manipulation beyond Coarse Synchronization}
\titlerunning{Lip-focused audio-visual consistency for speech manipulation}

\author{Jian Zong$^{1*}$, Chenghan Lin$^1$, Meng Liu$^1$, Zhao Xu$^1$, Junyu Wang$^1$, Siyuan Zhang$^1$, Peiyuan Jiang$^1$, Longbiao Wang$^1$, Jianwu Dang$^2$}

\authorrunning{Zong et al.}

\institute{
$^1$ Tianjin University, Tianjin, China\\
$^2$ Shenzhen Institutes of Advanced Technology, Chinese Academy of Sciences, Shenzhen, China
}

\begin{document}
\raggedbottom
\maketitle
\footnotetext[1]{Corresponding author. Email: zongjian\_66@tju.edu.cn}

\begin{abstract}
Speech-content manipulation changes what a visible person appears to say by replacing the speech track and, in stronger attacks, regenerating mouth motion to match the forged utterance. Such attacks are difficult to expose with texture artifacts or coarse lip-sync scores because modern retalking models explicitly optimize audio-visual synchrony. We test the narrower hypothesis that speech replacement still perturbs the local relationship between acoustic dynamics and mouth motion after speaker identity or coarse synchrony is controlled. To isolate this signal, we introduce SyncSpeech, an LRS3-derived benchmark containing real videos, audio-splicing cheapfakes and Wav2Lip retalking attacks under same-speaker and different-speaker replacement. We then introduce C-MIND, a lip-focused detector that aligns audio and mouth-motion representations at frame level and models their agreement and disagreement without transcripts, automatic speech recognition features or an unmanipulated reference. On SyncSpeech, pretrained SyncLip-C achieves 95.813\% scenario-averaged accuracy and 0.9943 AUC, outperforming sentence-level fusion, full-face inputs and a reproduced SyncNet baseline under the reported protocol. When trained on the corresponding FakeAVCeleb training split, C-MIND correctly classifies 278 of 280 samples in the selected unseen-speaker test subset, whereas accuracy falls to 82.775\% on unseen Video-Retalking attacks. These results identify lip-focused speech-articulation consistency as a useful but generator-sensitive forensic signal within the evaluated settings.
\end{abstract}

\section{Introduction}
Digital talking-head videos can now be edited at the level of words, voices and facial motion. Neural speech synthesis and voice conversion can generate or transform plausible utterances from limited speech evidence~\cite{kaur2023ttsreview,walczyna2023voiceconversion}, while text-based video editing, face reenactment, face swapping and neural rendering can synthesize convincing visual content~\cite{fried2019textbased,thies2016face2face,thies2019neuraltextures,nirkin2019fsgan}. Lip-synchronization systems further close the gap between manipulated speech and visible mouth motion~\cite{prajwal2020wav2lip,cheng2022videoretalking}. These tools support film production, dubbing and assistive communication, but they also enable attacks in which a public figure, financial authority or enrolled user appears to say something that was never spoken~\cite{kingra2023survey}.

The security-relevant case studied here is speech-content manipulation: the primary malicious objective is not to change who appears in the video, but to change the meaning of the spoken message. A weak attack can splice a different audio segment into the original video, whereas a stronger attack can regenerate the mouth region with Wav2Lip or Video-Retalking so that the forged speech remains visually synchronized. This setting weakens several common forensic cues. Synthesis artifacts can be generator-specific, while identity checks require a claimed or enrolled identity~\cite{korshunov2018speaker,agarwal2023watch,reiss2023deepfakes}. Low-level lip-sync offsets are exactly what retalking systems are trained to repair~\cite{prajwal2020wav2lip,cheng2022videoretalking}, weakening coarse synchronization as a forensic cue~\cite{chung2017syncnet,shahzad2023avlipsync}.

This motivates a narrower forensic hypothesis than transcript-level semantic verification. Speech production couples acoustic dynamics with visible articulation through timing, phonetic realization, co-articulation and speaker-specific motion patterns. Replacing speech may disturb this local coupling even when the modalities remain coarsely synchronized. Whole-face features can dilute the articulatory signal with pose, texture and expression, while sentence-level pooling can suppress short inconsistencies. The relevant observable is therefore frame-level audio--lip compatibility, not direct understanding of what was said.

We investigate this hypothesis with two contributions. First, SyncSpeech separates content replacement from speaker-identity mismatch across real, audio-splicing and retalking scenarios. Second, C-MIND compares aligned speech and mouth-motion embeddings through explicit frame-level interaction, with sentence-level, frame-fusion and full-face variants serving as controlled ablations. We evaluate seen and unseen speakers, one unseen retalking generator and within-dataset unseen-speaker generalization on a selected FakeAVCeleb subset~\cite{khalid2021fakeavceleb_dataset,khalid2021fakeavceleb}. Pretrained SyncLip-C reaches 95.813\% mean accuracy and 0.9943 AUC on SyncSpeech, but the lower Video-Retalking result shows that the evidence supports a bounded articulation-consistency claim rather than general deepfake state of the art.

\section{Related work and positioning}
Deepfake detection uses visual artifacts, identity consistency and cross-modal agreement~\cite{kingra2023survey}. Artifact-based systems search for synthesis traces in faces or textures, whereas speaker-inconsistency and claim-conditioned fact-checking methods test whether media agree with an asserted identity or other claimed facts~\cite{korshunov2018speaker,reiss2023deepfakes}. Word-conditioned facial-motion analysis instead relates recognized words to person-specific facial dynamics~\cite{agarwal2023watch}. These directions address important forgery cues, but they depend on generator traces, identity evidence or person-specific linguistic annotations that are not assumed here.

Audio-visual correspondence provides a reference-free alternative. SyncNet measures coarse temporal alignment between speech and visible mouth motion~\cite{chung2017syncnet}. Lip Sync Matters compares an observed lip sequence with one reconstructed by Wav2Lip~\cite{shahzad2022lipsync}, while AV-Lip-Sync+ combines self-supervised lip features with full-face visual features to expose multimodal inconsistency~\cite{shahzad2023avlipsync}. These methods establish that synchronization and pretrained speech representations are useful forensic cues, but reconstruction quality or coarse alignment can limit sensitivity once retalking has explicitly repaired lip sync.

Fine-grained approaches relate phonetic or linguistic content to visible articulation. Multi-phoneme selection learns phoneme-conditioned lip-shape detectors~\cite{lin2021multiphoneme}, and NPVForensics models correlations between non-critical phonemes and visemes~\cite{chen2023npv}. Lost in Translation instead compares independently recovered audio and visual transcripts~\cite{bohacek2024lost}, while word-conditioned facial-motion analysis requires person-specific word-motion evidence~\cite{agarwal2023watch}. C-MIND targets the same broad content-consistency problem but does not require phoneme labels, ASR transcripts, visual speech recognition or an enrolled person.

Self-supervised audio-visual models improve generalization by learning normal correspondence from real videos. Audio-visual anomaly detection uses an autoregressive model trained only on real data~\cite{feng2023selfsupervised}; AVFF learns masked cross-modal representations before supervised forgery classification~\cite{oorloff2024avff}; and recent synchronization frameworks jointly model within-modality and cross-modality temporal structure~\cite{anshul2025synchronization}. C-MIND is narrower than these general detectors: it isolates whether mouth-focused, aligned interaction remains informative when retalking has already repaired coarse synchrony.

Dataset shortcuts and generator shift complicate this interpretation. Recent work shows that audio-visual detectors can exploit spurious dataset cues such as leading silence, motivating real-only alignment and explicit shortcut analysis~\cite{smeu2025shortcuts}. SyncSpeech addresses a complementary control problem by separating same-speaker from different-speaker replacement and audio-only from lip-regenerated manipulation. These controls reduce obvious identity and synchronization shortcuts but cannot eliminate all generator or dataset artifacts. Accordingly, our strongest evidence comes from the within-benchmark ablations; broader comparison with recent self-supervised detectors and additional unseen generators remains necessary.

\section{Method}
\subsection{Problem formulation and overview}
C-MIND treats speech-content manipulation as a local audio--lip consistency problem. Given a synchronized acoustic sequence $\mathbf{x}_a$ and mouth-motion sequence $\mathbf{x}_v$, the model predicts whether the current audio-visual pair is real or manipulated without using transcripts, ASR-derived semantic features or an unmanipulated reference. The design is motivated by the observation that retalking can repair coarse lip synchrony while still perturbing the local relationship between acoustic dynamics and visible articulation.

\begin{figure}[H]
\centering
\includegraphics[width=0.9\textwidth]{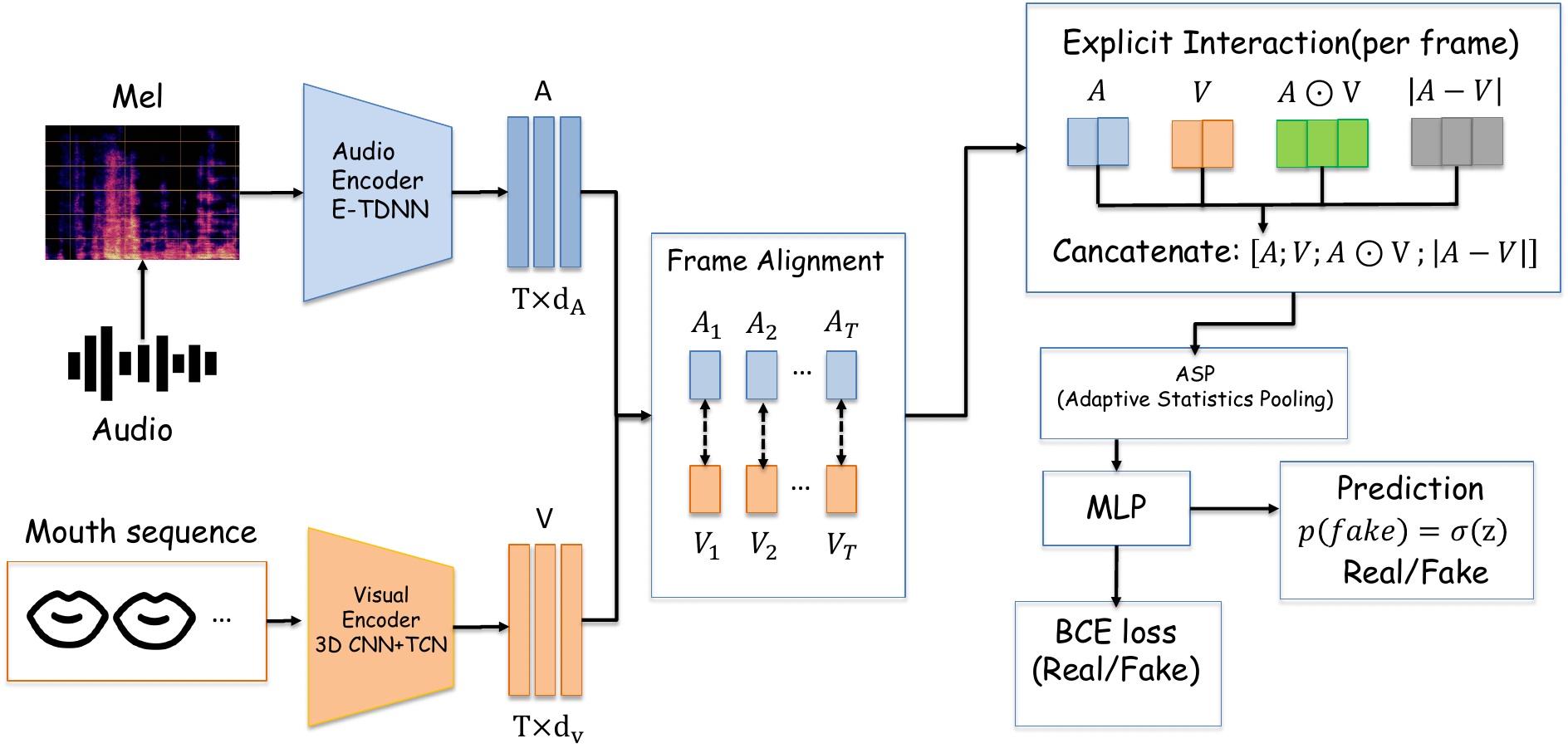}
\caption{\textbf{Overview of C-MIND (SyncLip-C).} Audio and visual encoders produce frame-level representations that are projected to a common dimension and temporally aligned. Their concatenation, element-wise product and absolute difference are pooled by ASP and classified by an MLP trained with binary cross-entropy.}
\label{fig:method}
\end{figure}

Figure~\ref{fig:method} summarizes the main SyncLip-C pipeline. Separate audio and visual encoders extract temporal representations, which are projected to a common channel dimension and aligned frame by frame. In the diagram, the alignment block denotes both temporal resampling and dimensional projection, yielding $\mathbf{A},\mathbf{V}\in\mathbb{R}^{T\times d}$ before element-wise interaction. At each time step, SyncLip-C combines the aligned embeddings using concatenation, element-wise product and absolute difference before temporal aggregation and classification. The controlled SyncLip-A, SyncLip-B and SyncLip-C variants are defined below but are not all depicted in the figure.

\subsection{Audio-lip encoding and temporal alignment}
The visual branch focuses on the mouth region rather than the full face to reduce variation from pose, identity texture and expression outside the articulators. It combines a 3D convolutional frontend, a residual image encoder and temporal convolution. The audio branch uses an extended time-delay neural network to encode frame-level acoustic dynamics. Because the two branches have different temporal strides and channel widths, a learned audio projection $P_a$ and visual interpolation operator $U_T$ map their outputs to a shared sequence length $T$ and dimension $d$:
\begin{equation}
\begin{aligned}
\mathbf{H}_a&=E_a(\mathbf{x}_a), & \mathbf{H}_v&=E_v(\mathbf{x}_v),\\
\mathbf{A}&=P_a(\mathbf{H}_a), & \mathbf{V}&=U_T(\mathbf{H}_v),
\qquad \mathbf{A},\mathbf{V}\in\mathbb{R}^{T\times d}.
\end{aligned}
\label{eq:aligned_embeddings}
\end{equation}
Here $\mathbf{a}_t$ and $\mathbf{v}_t$ denote the aligned audio and visual embeddings at time $t$. Retaining this alignment allows the detector to preserve short articulation inconsistencies that would be suppressed if both modalities were pooled before comparison.

\subsection{Frame-level cross-modal interaction}
SyncLip-C constructs an explicit interaction vector at every aligned time step:
\begin{equation}
\mathbf{c}_t=
[\,\mathbf{a}_t;\mathbf{v}_t;
\mathbf{a}_t\odot\mathbf{v}_t;
|\mathbf{a}_t-\mathbf{v}_t|\,],
\qquad t=1,\ldots,T,
\label{eq:interaction}
\end{equation}
where $[\,;\,]$ is channel-wise concatenation and $\odot$ is the Hadamard product. Concatenation retains modality-specific evidence, the product represents coordinate-wise agreement and the absolute difference represents disagreement. Adaptive statistics pooling, denoted by $\operatorname{ASP}$, converts a variable-length sequence into a fixed-dimensional representation, and an MLP $F_{\theta}$ produces the manipulation logit. The three decoder variants are
\begin{align}
z_A &= F_{\theta_A}\!\left(
[\operatorname{ASP}(\mathbf{H}_a);\operatorname{ASP}(\mathbf{H}_v)]
\right), \nonumber\\
z_B &= F_{\theta_B}\!\left(
\operatorname{ASP}(\{[\mathbf{a}_t;\mathbf{v}_t]\}_{t=1}^{T})
\right), \nonumber\\
z_C &= F_{\theta_C}\!\left(
\operatorname{ASP}(\{\mathbf{c}_t\}_{t=1}^{T})
\right),
\label{eq:variant_logits}
\end{align}
corresponding to sentence-level fusion, frame-level concatenation and frame-level explicit interaction. SyncFace-A and SyncFace-C use the same decoders as their SyncLip counterparts but replace the mouth-region input with the aligned full-face sequence. For every variant, $p(y=1\mid\mathbf{x}_a,\mathbf{x}_v)=\sigma(z)$, where $y=1$ denotes a manipulated video.

\subsection{Training objective}
The detector is optimized with binary cross-entropy. For a mini-batch of $N$ videos with labels $y_i\in\{0,1\}$ and logits $z_i$, the objective is
\begin{equation}
\mathcal{L}_{\mathrm{BCE}}
=-\frac{1}{N}\sum_{i=1}^{N}
\left[y_i\log\sigma(z_i)+(1-y_i)\log\bigl(1-\sigma(z_i)\bigr)\right].
\label{eq:bce_loss}
\end{equation}
The main pretrained configuration initializes both encoders from our separately trained audio-visual lip-motion speaker-verification model and trains newly initialized projection, interaction and classification layers. Section~\ref{sec:implementation} gives the preprocessing, optimization and checkpoint-selection details.

\section{Experiments}
\subsection{Datasets and evaluation scenarios}
We built SyncSpeech from LRS3 because it contains in-the-wild speech videos with diverse speakers and natural audio-visual variation~\cite{afouras2018lrs3}. The benchmark is designed to answer a specific forensic question: can a model detect speech-content replacement from the current audio-lip pair, even when the visible speaker and the lip motion appear plausible?

\begin{figure}[H]
\centering
\includegraphics[width=\textwidth]{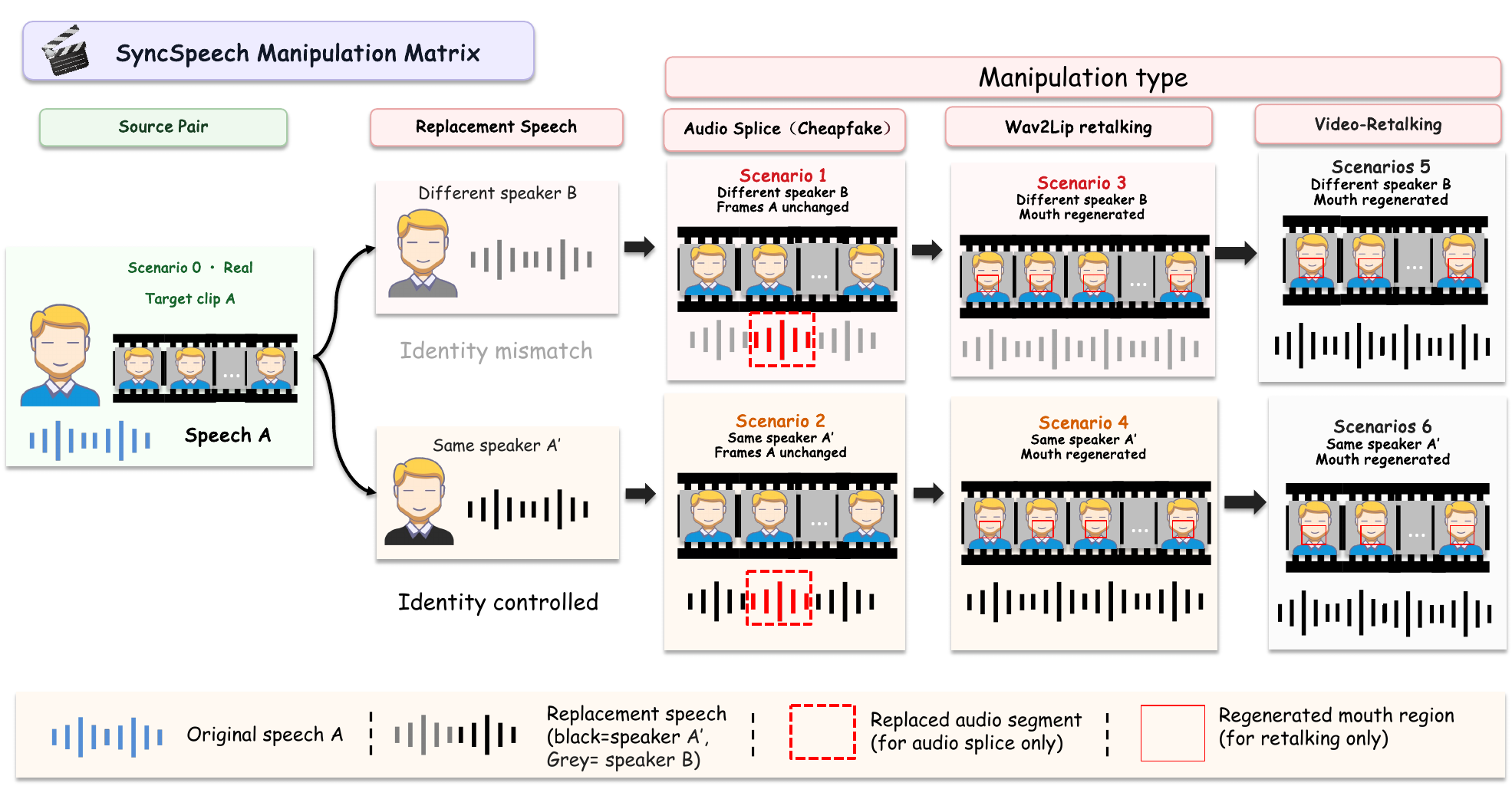}
\caption{\textbf{SyncSpeech manipulation matrix.} Scenario 0 supplies the real source pair. Different-speaker and same-speaker replacement speech produce the identity-mismatched and identity-controlled branches, respectively; audio splicing yields scenarios 1--2, Wav2Lip regeneration yields scenarios 3--4 and Video-Retalking provides the unseen-generator scenarios 5--6.}
\label{fig:syncspeech_protocol}
\end{figure}

SyncSpeech contains five in-distribution scenarios (0--4). Scenario 0 contains real videos. Scenario 1 contains cheapfake heterologous attacks, in which speech from a different speaker is spliced into the target video. Scenario 2 contains cheapfake homologous attacks, in which speech from the same speaker is spliced into a different target segment. Scenario 3 contains Wav2Lip heterologous attacks and scenario 4 contains Wav2Lip homologous attacks~\cite{prajwal2020wav2lip}. The homologous cases are deliberately difficult because the audio identity is consistent with the target speaker, leaving speech-articulation compatibility as the main observable evidence. To test generator generalization, we additionally construct two unseen-generator Video-Retalking scenarios: scenario 5 uses the heterologous setting and scenario 6 uses the homologous setting~\cite{cheng2022videoretalking}. Figure~\ref{fig:syncspeech_protocol} summarizes the complete construction and separates the identity-mismatched and identity-controlled paths.

The training split contains 20,000 real and 20,000 fake videos. The SyncSpeech test split contains 320 real and 1,280 fake videos, equally divided between seen and unseen speaker settings. The Video-Retalking test set contains 627 generated fake videos, corresponding to approximately 98\% of the intended generated samples after failed generations were excluded. For the external FakeAVCeleb benchmark~\cite{khalid2021fakeavceleb_dataset,khalid2021fakeavceleb}, C-MIND is trained on the corresponding 8,255 real and 8,268 fake training samples and evaluated on a selected unseen-speaker test subset containing 70 real and 210 fake samples (Table~\ref{tab:data}). This experiment measures within-dataset unseen-speaker generalization rather than zero-shot cross-dataset transfer. Clip durations range from 0.54 to 6.98 s, so many samples provide only a short temporal window for detecting manipulation.

\subsection{Manipulation generation and split controls}
SyncSpeech was constructed from LRS3 by pairing each target video with replacement speech whose content differed from the target segment. The replacement speaker was either the same as the target speaker (homologous) or a different speaker (heterologous). Gender was balanced during pairing, problematic speaker labels were manually avoided and the train/test split followed the speaker setting used for seen- and unseen-speaker evaluation. When necessary, a speed factor between 0.8 and 1.2 adjusted the replacement utterance to the target duration.

Cheapfake samples replace only the target audio track. Wav2Lip samples regenerate the visible mouth region from the target video and replacement audio~\cite{prajwal2020wav2lip}, while Video-Retalking samples use the corresponding retalking setting to test transfer to an unseen generator~\cite{cheng2022videoretalking}. Failed outputs caused by face tracking or synthesis were excluded; the retained Video-Retalking set contains 627 videos, approximately 98\% of the intended generated set. Table~\ref{tab:data} reports the final evaluation splits.

\begin{table}[H]
\centering
\caption{\textbf{Dataset composition.} Seen and unseen indicate whether the speaker appears in the training split.}
\label{tab:data}
\small
\renewcommand{\arraystretch}{1.08}
\setlength{\tabcolsep}{9pt}
\begin{tabular}{lllr}
\toprule
Dataset & Subset & Label or condition & Videos \\
\midrule
\multirow{6}{*}{SyncSpeech} & Train & Real & 20,000 \\
& Train & Fake & 20,000 \\
& Test & Real, seen speakers & 160 \\
& Test & Real, unseen speakers & 160 \\
& Test & Fake, seen speakers & 640 \\
& Test & Fake, unseen speakers & 640 \\
\addlinespace[1.5pt]
\multirow{2}{*}{Video-Retalking} & Test & Fake, seen speakers & 313 \\
& Test & Fake, unseen speakers & 314 \\
\addlinespace[1.5pt]
\multirow{4}{*}{FakeAVCeleb} & Train & Real & 8,255 \\
& Train & Fake & 8,268 \\
& Test & Real, unseen speakers & 70 \\
& Test & Fake, unseen speakers & 210 \\
\bottomrule
\end{tabular}
\end{table}

\subsection{Implementation details}
\label{sec:implementation}
For each frame, facial landmarks were detected and used to align the face to a reference shape. Videos were sampled at 25 frames per second, and each training or evaluation example used a 2~s clip. A fixed mouth-region crop was resized to 96$\times$96 and converted to grayscale. Audio was resampled to 16~kHz and represented by 80-bin log-mel features with a 25~ms analysis window and a 10~ms hop. The audio features were synchronized with the visual sequence before cross-modal comparison. Samples with failed face detection or unusable generation outputs were excluded only from the affected generated subsets.

The pretrained SyncLip-C encoders were initialized from our separately trained audio-visual lip-motion speaker-verification model. Its visual stage used 29-frame mouth segments with 88$\times$88 random crops and horizontal flipping. Its audio stage used MFCC features and an AM-Softmax objective. Because the final detector uses 80-bin log-mel features rather than the MFCC input used during pretraining, the acoustic input projection was newly initialized and pretrained audio layers were loaded only where tensor dimensions matched.

For forgery classification, newly initialized projection, interaction and classification layers used AdamW~\cite{loshchilov2019decoupled} with an initial learning rate of $10^{-4}$. Pretrained backbone layers used $3\times10^{-5}$ only in fine-tuning ablations where they were unfrozen. We used cosine learning-rate decay with 10\% warmup, weight decay of $10^{-4}$, a global batch size of 32 and at most 30 epochs. The main configuration froze the pretrained backbone; partial fine-tuning of its last two blocks was used only for ablation.

\subsection{Baselines, metrics and model selection}
We compare sentence-level lip and face fusion (SyncLip-A and SyncFace-A), frame-level concatenation (SyncLip-B), explicit frame interaction with lip or face inputs (SyncLip-C and SyncFace-C), and pretrained SyncLip-C. SyncNet~\cite{chung2017syncnet} is reproduced as a low-level lip-sync baseline, and published AV-Lip-Sync results~\cite{shahzad2022lipsync} are used for the selected FakeAVCeleb comparison. These comparisons test three questions: whether the mouth region is more informative than the full face, whether frame-level evidence is necessary and whether explicit interaction and pretraining improve detection.

A video-level validation subset was held out from training, and no clip from a validation or test video was used for training. The final checkpoint was selected solely by validation AUC. The probability threshold maximizing balanced accuracy on validation data was then fixed for test evaluation. We report scenario-wise accuracy and AUC, separate seen- and unseen-speaker results, and Wilson confidence intervals where the denominator is recoverable. FakeAVCeleb is treated as supporting within-dataset unseen-speaker evidence because C-MIND is trained on its corresponding training split, while Video-Retalking is the primary unseen-generator test.

\subsection{Human observer protocol}
Ten participants aged 21--27 years judged whether SyncSpeech videos were real or manipulated. The group was gender balanced and had high English proficiency. No participant-identifying information is reported, and responses are analyzed only at aggregate scenario level. This comparison provides descriptive context for perceptual difficulty; it is not used for model selection, threshold calibration or hypothesis testing.

\section{Results and Discussion}
\subsection{Ablation validates lip-focused frame-level interaction}
We first test whether the evidence is carried by the mouth region and whether frame-level modeling is necessary. Sentence-level full-face fusion performs poorly, with 31.688\% mean accuracy and 0.5075 AUC (Table~\ref{tab:syncspeech_ablation}). Replacing the full face with the lip region improves detection of Wav2Lip attacks but still fails on audio-splicing cheapfakes, showing that sentence-level fusion loses local temporal evidence.

Frame-level models change the result. SyncLip-B reaches 92.375\% mean accuracy and 0.9885 AUC. SyncLip-C, which explicitly models product and difference interactions between aligned modalities, reaches 94.313\% mean accuracy and 0.9872 AUC when trained from scratch. The full-face interaction model, SyncFace-C, remains weak (56.000\% mean accuracy and 0.5297 AUC), showing that full-face inputs are not beneficial under the reported protocol. With audio-visual speaker-verification pretraining, SyncLip-C reaches 95.813\% mean accuracy and 0.9943 AUC. The difference between SyncLip-B and scratch SyncLip-C is modest in AUC, so the main supported conclusion is that frame-level lip-focused comparison is essential, while explicit product/difference interactions and pretraining provide additional gains in this evidence chain.

\begin{table}[H]
\centering
\caption{\textbf{Ablation on SyncSpeech.} Scenario accuracy (\%), its five-scenario arithmetic mean and all-scenario AUC. Bold marks the best result in each row, including ties.}
\label{tab:syncspeech_ablation}
\small
\renewcommand{\arraystretch}{1.08}
\setlength{\tabcolsep}{6.5pt}
\begin{tabular}{lrrrrrr}
\toprule
Metric & Face-A & Lip-A & Lip-B & Lip-C & Face-C & Lip-C (pre.) \\
\midrule
Scenario 0 & 81.250 & 93.125 & \textbf{99.375} & 98.125 & 45.938 & 99.063 \\
Scenario 1 & 18.125 & 5.938 & 91.563 & 91.875 & 58.750 & \textbf{95.625} \\
Scenario 2 & 19.375 & 6.563 & 93.750 & 93.750 & 59.375 & \textbf{95.938} \\
Scenario 3 & 19.063 & 83.438 & 89.375 & \textbf{95.000} & 58.125 & \textbf{95.000} \\
Scenario 4 & 20.625 & 82.188 & 87.813 & 92.813 & 57.813 & \textbf{93.438} \\
\addlinespace[1.5pt]
Mean accuracy & 31.688 & 54.250 & 92.375 & 94.313 & 56.000 & \textbf{95.813} \\
AUC & 0.5075 & 0.7410 & 0.9885 & 0.9872 & 0.5297 & \textbf{0.9943} \\
\bottomrule
\end{tabular}
\end{table}

\subsection{Generalization across speakers, datasets and generators}
The pretrained SyncLip-C model maintains strong performance for both seen and unseen speakers (Table~\ref{tab:pretrained}). The mean accuracy is 97.250\% for seen speakers and 94.375\% for unseen speakers. The drop is expected because unseen speakers introduce new articulation styles, but the AUC remains high at 0.9918. Homologous attacks are harder than heterologous attacks because the audio identity is no longer an obvious cue; nevertheless, the model achieves 95.938\% accuracy on cheapfake homologous attacks and 93.438\% accuracy on Wav2Lip homologous attacks.

{\hbadness=10000 The scratch-to-pretrained comparison tests whether prior audio-visual speaker-articulation structure benefits the detector. Accuracy rises from 94.313\% to 95.813\%, supporting pretrained initialization under this protocol. Because upstream speaker verification is a different task, we do not count it as an additional forgery-detection robustness result.\par}

\begin{table}[H]
\centering
\caption{\textbf{Seen and unseen speaker results for pretrained SyncLip-C on SyncSpeech.}}
\label{tab:pretrained}
\small
\renewcommand{\arraystretch}{1.08}
\setlength{\tabcolsep}{10pt}
\begin{tabular}{lrrr}
\toprule
Scenario & Seen & Unseen & Mean \\
\midrule
0 Real & 98.750 & 99.375 & 99.063 \\
1 Cheapfake, heterologous & 97.500 & 93.750 & 95.625 \\
2 Cheapfake, homologous & 99.375 & 92.500 & 95.938 \\
3 Wav2Lip, heterologous & 95.000 & 95.000 & 95.000 \\
4 Wav2Lip, homologous & 95.625 & 91.250 & 93.438 \\
\addlinespace[1.5pt]
Mean accuracy & 97.250 & 94.375 & \textbf{95.813} \\
AUC & 0.9963 & 0.9918 & \textbf{0.9943} \\
\bottomrule
\end{tabular}
\end{table}

We next compare C-MIND with external baselines (Table~\ref{tab:external}). After training on the corresponding FakeAVCeleb training split, C-MIND correctly classifies 278 of 280 samples in the selected unseen-speaker lip-sync forgery test subset. The original AV-Lip-Sync paper reports 98\%, 97\% and 96\% accuracy on its separate 140-sample Faceswap+Wav2Lip, FSGAN+Wav2Lip and Wav2Lip test sets, respectively~\cite{shahzad2022lipsync}. Because the results use different sample constructions, they provide contextual rather than strictly paired comparison; these small, saturated subsets are supporting external-benchmark evidence rather than the main robustness result.

\begin{table}[H]
\centering
\caption{\textbf{Comparison with external baselines.} AV-Lip-Sync values retain source precision~\cite{shahzad2022lipsync}; SyncSpeech + Video-Retalking averages are arithmetic means of the four scenario groups.}
\label{tab:external}
\small
\renewcommand{\arraystretch}{1.08}
\setlength{\tabcolsep}{7pt}
\begin{tabular}{llr}
\toprule
Dataset & Method or scenario & Accuracy (\%) \\
\midrule
\multirow{4}{*}{FakeAVCeleb} & AV-Lip-Sync, Faceswap+Wav2Lip & 98 \\
& AV-Lip-Sync, FSGAN+Wav2Lip & 97 \\
& AV-Lip-Sync, Wav2Lip & 96 \\
& C-MIND / SyncLip-C, average & 99.286 \\
\addlinespace[1.5pt]
\multirow{5}{*}{\shortstack{SyncSpeech +\\Video-Retalking}} & SyncNet, scenario 0 & 25.625 \\
& SyncNet, scenarios 1--2 & 79.063 \\
& SyncNet, scenarios 3--4 & 79.688 \\
& SyncNet, scenarios 5--6 & 80.223 \\
& SyncNet, average & \textbf{66.150} \\
\addlinespace[1.5pt]
\multirow{5}{*}{\shortstack{SyncSpeech +\\Video-Retalking}} & C-MIND, scenario 0 & 99.063 \\
& C-MIND, scenarios 1--2 & 95.782 \\
& C-MIND, scenarios 3--4 & 94.219 \\
& C-MIND, scenarios 5--6 & 82.775 \\
& C-MIND, average & \textbf{92.960} \\
\bottomrule
\end{tabular}
\end{table}

On the combined SyncSpeech and Video-Retalking evaluation, C-MIND reaches 92.960\% group-averaged accuracy, whereas the reproduced SyncNet baseline reaches 66.150\%. The largest gap occurs for real videos, where SyncNet rejects many genuine samples. On unseen Video-Retalking attacks, however, C-MIND drops to 82.775\%, corresponding to 519 correct classifications among 627 generated videos and only moderately above SyncNet's 80.223\%. This result does not show that generator shift is solved; it instead shows better calibration across real and fake videos under the reported protocol while confirming that cross-generator retalking remains a limitation.

\subsection{Exploratory human-observer comparison}
We include an aggregate human-observer comparison only as descriptive context for the perceptual difficulty of SyncSpeech. Ten participants aged 21--27 years, with balanced gender composition and high English proficiency, judged videos from the SyncSpeech scenarios. Human observers performed well on real videos, reaching 91.56\% accuracy on scenario 0, and also identified audio-splicing cheapfakes with 86.56\% accuracy on heterologous attacks and 88.44\% on homologous attacks. Their accuracy dropped substantially on Wav2Lip retalking attacks, reaching 37.19\% on heterologous deepfakes and 35.00\% on homologous deepfakes. The average human accuracy across scenarios was 67.750\%. This reader study is not used for model selection, threshold calibration or the primary statistical claims of the paper; it is retained to indicate that the retalking samples are not always obvious to unassisted viewers.

The low human accuracy on scenarios 3 and 4 indicates that the Wav2Lip attacks are not always obvious visual failures for unassisted observers. In contrast, pretrained SyncLip-C reaches 95.813\% mean accuracy on SyncSpeech, including 95.000\% on Wav2Lip heterologous attacks and 93.438\% on Wav2Lip homologous attacks. This gap is a qualitative motivation for automated audio-lip analysis rather than independent proof of superiority over human judgement. We interpret the model evidence as speech-articulation inconsistency rather than direct linguistic semantic understanding, because the detector does not use transcripts, ASR-derived semantic representations or an unmanipulated reference video.

\begin{table}[htbp]
\centering
\caption{\textbf{Exploratory human-observer comparison on SyncSpeech.} Human observers are accurate on real videos and audio-splicing cheapfakes but struggle with lip-synchronized speech-content replacement.}
\label{tab:human}
\renewcommand{\arraystretch}{1.08}
\setlength{\tabcolsep}{11pt}
\begin{tabular}{lrr}
\toprule
Scenario & Human (\%) & SyncLip-C (\%) \\
\midrule
0 Real & 91.560 & 99.063 \\
1 Cheapfake, heterologous & 86.560 & 95.625 \\
2 Cheapfake, homologous & 88.440 & 95.938 \\
3 Wav2Lip, heterologous & 37.190 & 95.000 \\
4 Wav2Lip, homologous & 35.000 & 93.438 \\
\addlinespace[1.5pt]
\textbf{Average across scenarios} & \textbf{67.750} & \textbf{95.813} \\
\bottomrule
\end{tabular}
\end{table}

\subsection{Uncertainty and scope of the reported results}
The reported numbers are fixed-test-set estimates under the stated split and model-selection procedure, not distribution-free robustness guarantees. For pretrained SyncLip-C on SyncSpeech, 95.813\% accuracy over 1,600 videos gives a Wilson 95\% confidence interval of 94.72--96.69\%. For Video-Retalking scenarios 5--6, 82.775\% accuracy over 627 videos gives an interval of 79.62--85.53\%. These intervals quantify sampling uncertainty for the evaluated sets but do not replace per-seed dispersion, bootstrap AUC intervals or stress tests under compression and audio noise.

\subsection{Interpretation, limitations and scope}
The experiments support one central conclusion: within SyncSpeech, the most informative tested representation is a frame-level comparison between speech and mouth dynamics. Audio-only cheapfakes remove visual synthesis artifacts, same-speaker attacks reduce identity mismatch and retalking reduces coarse synchronization error. Across these controls, lip-region models outperform full-face counterparts and frame-level models outperform sentence-level fusion. The result therefore supports local articulation compatibility more directly than generic face appearance or global audio-visual matching.

This conclusion is intentionally mechanistic rather than semantic. C-MIND does not use transcripts, ASR features or an unmanipulated reference, so its decisions may combine imperfect co-articulation, speaker-specific motion and generator-dependent mouth dynamics. The observer study adds perceptual context---participants often accepted Wav2Lip attacks as real---but it is not an independent test of model superiority. Likewise, strong selected-subset performance after training on FakeAVCeleb is supporting external-benchmark evidence rather than a zero-shot cross-dataset claim.

Generator shift is the clearest limitation. Accuracy falls to 82.775\% on Video-Retalking scenarios 5--6, only moderately above the reproduced SyncNet result of 80.223\%. The study also lacks compression, codec, resolution and audio-noise stress tests, and its same-protocol baselines do not cover the strongest recent self-supervised audio-visual detectors. Thus, the present work establishes a controlled benchmark and a useful articulation-level signal, while broader generator, domain and degradation robustness remain the necessary next tests.

\section{Conclusion}
This paper frames speech-content manipulation as a local audio--lip consistency problem. SyncSpeech separates audio-only from lip-regenerated attacks and controls whether replacement speech comes from the same or a different speaker. C-MIND compares aligned speech and mouth-motion representations rather than whole-face appearance or sentence-level fusion. Pretrained SyncLip-C reaches 95.813\% scenario-averaged accuracy and 0.9943 AUC on SyncSpeech, remains effective for unseen speakers and achieves 99.286\% accuracy after separate training on FakeAVCeleb and evaluation on its selected unseen-speaker subset. Together, these results establish articulation-level consistency as a useful forensic signal when texture artifacts, identity mismatch and coarse synchronization are weakened as shortcuts.

The signal is not yet generator-invariant. Accuracy decreases to 82.775\% on the unseen Video-Retalking attacks, and the present evaluation does not cover additional generators, compression, codec changes, resolution shifts or audio noise. Future work should therefore combine articulation-focused modeling with broader self-supervised audio-visual representations and evaluate it under multi-generator, cross-domain and real-world degradation protocols.

\begingroup
\hbadness=10000
\bibliographystyle{splncs04}
\bibliography{references}
\endgroup

\end{document}